\documentclass[modern,twocolumn]{aastex62}
                \usepackage{float}
                \usepackage{times}        
                \usepackage{graphicx}

                        \def\spose#1{\hbox to 0pt{#1\hss}}         
                        \def\simlt{\mathrel{\spose{\lower 3pt\hbox{$\mathchar"218$}}         
                        \raise 2.0pt\hbox{$\mathchar"13C$}}} 
                        \def\simgt{\mathrel{\spose{\lower 3pt\hbox{$\mathchar"218$}} 
                        \raise 2.0pt\hbox{$\mathchar"13E$}}} 

                \received{}
                \revised{}
                \accepted{}
                \submitjournal{ApJ}
                
                \shorttitle{Local metallicity distribution function}
                \shortauthors{Mishurov \& Tkachenko}
                
\begin{document}
                
                \title{Local metallicity distribution function derived from Galactic large~-~scale radial iron pattern modelling}
                
                \correspondingauthor{Yu.N.Mishurov}
                \email{unmishurov@sfedu.ru}

                \author[0000-0002-8068-5139]{Yu.N.Mishurov}
                \affiliation{Physical Faculty, Southern Federal University, 5 Zorge, Rostov-on-Don, 344090, Russia}
                
                \author[0000-0003-2695-864X]{R.V.Tkachenko}
                \affiliation{Physical Faculty, Southern Federal University, 5 Zorge, Rostov-on-Don, 344090, Russia}

                \begin{abstract}
                We develop an approach for fitting the results of modeling of wriggling radial large scale iron pattern along the Galactic disk, derived over young (high massive) Cepheids, with the metallicity distribution, obtained using low mass long living dwarf stars in the close solar vicinity. For this, at the step of computing of the theoretical abundance distribution over low mass stars in the solar vicinity we propose to redefine the initial mass function so as the resulting theoretical stellar distribution over masses would be close to the distribution in the observed sample. By means of the above algorithm and subsequent corrections of the theoretical metallicity distribution function, described in literature, we have achieved fairly well agreement of the theoretical and observed metallicity distribution functions for low mass stars in the local solar vicinity.
                                
                \end{abstract}
                

                \section{Introduction} \label{sec:intro}
                
                It has been recognized long ago that, studies of the Galaxy enrichment by heavy elements play an important role in understanding of the details of its evolution. Along with kinematical and dynamical approaches, abundances of heavy elements open an opportunity to reconstruct more complete picture of the Galaxy history.
                 
                Early simulations of the Galactic chemical evolution were restricted by the local solar vicinity since at that time the observational data on abundances were only accessible for low mass dwarf stars situated close to the Sun. It is worthwhile to notice, that in the local vicinity such stars are large in number and their abundances are determined with good accuracy. Thereto, they have long life times\footnote{Life time of a star is the time period from the star birth to its `death'.} which can be comparable (or even greater) with the Galactic age. Due to that these stars provide us with information about the Galactic history since almost its origins. 
                
                What type of information do such studies bring the researchers? The observations enable to estimate the mean metallicity in the solar vicinity and the metallicity distribution function (MDF) over the dwarf stars. Therefore, any theory has to interpret the observational data. However, since the first studies the researchers faced with a problem~--~ the deficiency of metal~-~poor stars in observations comparing with their number predicted by a theory. This divergence of theory with observations was called as `G~-~dwarfs' problem (van den Bergh 1962).
                                
                To improve the convergence of the theory with observations several modifications were proposed, for instance, the so~-~called `inside~-~out' model of the Galactic disk formation first proposed by Larson (1976) and developed in a series of papers by Matteucci \& Fran\c{c}ois (1989), Chiappini et al. (1997, 2001), Fran\c{c}ois et al. (2004) and others. According to this scenario a convergence between the theoretical and observed MDFs can be reached if the time scale, $t_f$, of low abundant gas infall onto the Galactic disk depends on the Galactocentric distance, $r$, so as at the solar distance the time scale is long $t_f\sim7$ Gyr. Due to that in the past the rate of low metal stars birth was decreased in the solar vicinity and they do not contribute significantly to the metal~-~poor tail of the observed MDF.
                
                The possible dependences of $t_f$ on $r$ were analyzed in details by Moll\'a, D\'iaz, Gibson, Cavichia \& L\'opez ~-~S\'anchez (2016). Using the derived representation for $t_f(r)$ Moll\'a, D\'iaz, Yago \& Gibson (2017) develop a model for the Galactic disk enrichment by heavy elements. In addition, the authors take into account such subtle effects like the interstellar gas conversion from neutral phase to molecular one with subsequent star formation. According to the cited papers the theoretical MDFs coincides with the observed one well.
                
                Another idea for solution the problem was developed by Sommer~-~Larsen (1991, hereafter SL91) and Wyse \& Gilmore (1995) and others. The authors demonstrate that the deficiency of low metallicity stars (in observations) can be associated with stellar migration. That is why dwarf stars of intermediate metallicities are under~-~represented in the solar vicinity. The authors propose some correction factors which enable to improve the agreement of theory with observations.
                                
                This approach was elaborated further in a series of papers by Haywood (2001; 2002; 2006; 2014) as well by Haywood, Snaith, Lehnert, Matteo \& Khoperskov (2019) where they notice that several assumptions, introduced in the inside~-~out model, are not realized. In particular, the authors state that stars, seen in the solar vicinity, are only the `tip of the iceberg'; the modeling of the Galactic disk by means of independent rings, which do not interact, is insufficient and the long time scale for infall gas is not necessary. They show that the G~-~dwarfs problem can be solved even in the closed~-~box model if the stellar wandering are taken into account.\footnote{The stellar mixing between thin and thick disks results in appearance in the solar vicinity stars from the thick disk which where rapidly enriched by low abundant matter at early times.} 
                
                Here we would like to emphasize that the observational studies  of abundances in dwarf stars provide us with information about the Galactic disk evolution in a very small region of the disk close to the Sun within $\sim100$ pc.

                After appearance of new data on abundances in bright objects, observed at large distances from the Sun, the problem of the Galactic chemical evolution has moved into another plane~--~to explain the Galactic radial gradients of chemical elements (e.g., Wyse \& Silk 1985; Portinari \& Chiosi 1999; Cescutti, Matteucci, Fran\c{c}ois \& Chiappini 2007; Moll\'a et al. 2017, etc.).
                
                Very interesting information on the Galactic large scale abundance distributions was provided with Cepheids~--~bright young stars which have sufficiently precisely determined distances. An intrigue feature of the Galactic radial abundance pattern, demonstrated by Cepheids, proved the unexpected wriggling metallicity radial distribution with the bend in the abundances located close to the solar Galactocentric distance (see the series of papers by Andrievsky et al. 2002 and references therein). For explanation of this nontrivial structure Mishurov, L\'epine \& Acharova (2002) proposed the idea of combined effect of the corotation resonance, located close to the Sun (e.g., Marochnik, Mishurov \& Suchkov 1972; Cr\'ez\'e \& Mennesier 1973; L\'epine, Mishurov \& Dedikov 2001; Dias, Monteiro, L\'epine \& Barrows 2019 and references therein) and the turbulent diffusion of metals in the interstellar medium on the formation of the fine structure in iron abundance pattern. This idea was elaborated in the subsequent papers by Acharova, L\'epine \& Mishurov (2005), Acharova, Mishurov \& Rasulova (2011, AMR11), Acharova, Mishurov \& Kovtyukh (2012, hereafter AMK12), Acharova, Gibson, Mishurov \& Kovtyukh (2013, hereafter AGMK13), etc. Perhaps the most difficult problem was to explain the similar radial patterns of oxygen and iron since their sources~--~core collapse supernovae and supernovae Type Ia~--~have different nature.
                
                In Mishurov \& Tkachenko (2018, 2019; hereafter MT18 and MT19, respectively) we revised the model by AGMK13 using new more precise and large in number observations of abundances and some other refinements (details see in the cited papers). Our theoretical radial large scale patterns of oxygen and iron coincide with the observed ones very good. It is worthwhile to notice that, in the papers we come to a conclusion that the model with rapid infall rate $t_f\sim2$ Gyr, independent on the Galactic radius $r$, describes the radial iron distribution very good.\footnote{Despite the result of our statistical analysis that the constant and small $t_f$ is preferable comparing to the one dependent on $r$, we do not believe that the inside~-~out scenario should be rejected. Perhaps, the observational data are insufficient to take into account the radial dependence of $t_f$.}

                In conclusion, we emphasize that Cepheids are very young objects (according to AMK12 their ages do not exceed significantly $\sim100$ Myr). Hence they give the abundance patterns within a wide Galactocentric radial range (from about 5 till 14 kpc) but at present epoch. Unlike that, low luminous dwarfs provide us with information about the disk temporal evolution since its origins but in the local solar vicinity.

                The aim of the present paper is to answer the question: can the theory by MT19, elaborated for explanation of the large scale iron radial distribution in the Galactic disk using young Cepheids, be compatible with the iron metallicity distribution observed over the long lived dwarfs within the local solar vicinity?

                                
                                \section{Basic Ideas and Methods}

                To feel deeply the problem we remind in short the models developed in MT19 for the iron abundance pattern formation along the Galactic disk (in what follows we mainly examine the processes in the Galactic thin disk). Unlike the traditional modeling of the Galactic disk abundance patterns, in the above cited papers we focus on reconstruction of the wriggling radial iron distribution.
                
                \subsection{Formation of the large scale wriggling iron radial pattern}
                
                The central idea for explanation of the above feature is to take into account the stimulating effect of spiral arms on the disk enrichment, especially the effect of the corotation resonance. \footnote{The corotation resonance is located in a ring like region within the Galactic plane, which center coincides with the one for the disk. Here the rotation velocity of the Galactic matter, $\Omega(r)$, is equal to the one for spiral density waves, $\Omega_P$, responsible for arms. The Galactocentric radius, $r_c$, of the corotation ring is determined by the following equality: $\Omega(r_c)=\Omega_P$.} 
                The idea is as follows: metals are provided to the Galactic disk with their sources. Hence spiral arms may influence on the rate of enrichment the Galactic disk if the sources of the elements are concentrated in spiral arms where they were born. For this the sources have to be short living objects: as it was shown in AGMK13 their life times, $\tau_m$, have to satisfy the following inequality: $\tau_m\simlt\tau_b\sim100$ Myr. Iron is supplied to the disk by the three sources: core collapse supernovae (CC~SNe), prompt and tardy supernovae Type Ia (SNeIa~-~P and SNeIa~-~T, respectively). Life times of CC~SNe do not exceed $\sim30$ Myr. According to Maoz, Keren \& Gal-Yam (2010), Li et al. (2011) and others the upper value for $\tau_m$ of SNeIa~-~P is estimated to be $\simlt400$ Myr. Following the above restriction, as the short living part of SNeIa~-~P we consider only those for which $\tau_m\simlt100$ Myr (see also Bonaparte et al. 2013). The remaining part of SNeIa~-~P (for which $\tau_m>100$ Myr) as well as SNeIa~-~T we refer to long living objects supposing that their contribution to the disk enrichment is not subjected to the influence of spiral arms. We denote them as SNeIa~-~T.
                
                Such division of iron sources into two groups results in two types of representations of SFR: for short living (i.e. high mass) stars, $\psi_H$, and long living (low mass) objects, $\psi_L$. According to AGMK13 the representations for the SFRs are as follows:
                                  \begin{equation}
                                        \psi=\left\{
                                \begin{array}{ll}
                                \psi_H(r,t)=\beta|\Delta\Omega|\mu_g^{1.5}, & \mbox{if  } \tau_m\leq \tau_b, \\        
                        \psi_L(r,t)=\nu\mu_g^{1.5}\,\,\,\,\,\,\, ,& \mbox{if  } \tau_m > \tau_b,
                                \end{array} \right.
                                \end{equation}
                where $\mu_g(r,t)$ is the surface density of the total (atomic + molecular) gaseous Galactic component, the life time of a star, $\tau_m$, is dependent on stellar mass $m$ (all masses are in solar units) by the following relation:
                \begin{equation} 
                log(\tau_m)=1.056-3.8 log(m)+log(m)^2, 
                \end{equation}
                $\beta$ and $\nu$ are norming constants, $\Delta\Omega= \Omega(r)-\Omega_P$. For the rotation curve the one from Clemens (1985) was adopted in MT18 and MT19. According to AMR11 the corotation resonance is located at $r_c=7$ kpc (the Sun is located at $r_{\odot}=7.9$ kpc). Equating $\Omega_P=\Omega(r_c)$ we derive the rotation velocity of the spiral pattern. 
        
        The above explicit division of the representation for SFRs needs some more detailed explanations. First of all, remind that in the traditional theories for the Galactic disk nucleosynthesis only the radial distributions of metallicity are studied. Thereto the bulk of large scale observational distributions for heavy elements in the Galactic disk are mainly available as the radial ones. These imply that equations, used in the theoretical modelling for the Galactic disk enrichment, are averaged over the Galactic azimuth angle. The corresponding procedure for averaging of the equations was described in details by AGMK13 (see Section 2.3 therein). This azimuth averaging explains why in our model the density of short living objects depends only on the Galactocentric distance $r$ despite the visible young bright objects in galaxies are distributed within thin strings located along spiral arms.
        
        At first sight, perhaps the more general representation for SFR would be a linear superposition of $\psi_H$ and $\psi_L$ (such idea was realized in Andrievsky et al. 2004). However in MT18 and MT19 (see also AGMK13) we prefer to separate explicitly the SFRs for the short and long living objects. The fact is, according to Mishurov \& Acharova (2011) the long living stars will be scattered over a significant portion of the Galactic disk under the influence of the Galactic spiral arms and they will not keep in their memory that they were born in spiral arms. On the other hand, the nonmonotone radial distribution of short living objects is extremely important for explanation of the wriggling iron Galactic radial pattern.
        
        The crucial point is: Is it possible to observe the gap in the radial distribution of short living SNe predicted by equation (1)? Recently Karapetyan et al. (2018) have studied 269 galaxies and shown that near the corotation resonance there is a valley in the radial distribution of short living CC~SNe (see the inset in their figure~6). Hence our prediction that there should be a gap in the radial distribution of short living objects (like, e.g. CC~SNe) has gotten observational support.
                
                Mathematically the enrichment rate of the disk by iron is represented as follows: 
                $$E_{Fe} = P^{\rm cc}_{Fe}R^{\rm cc}+P^{\rm P}_{Fe}R^{\rm P} + P^{\rm T}_{Fe} R^{\rm T},$$ 
                where $P^{\rm cc,\,P,\,T}_{Fe}$ are the mean iron masses ejected per CC~SNe, short living (in our definition) SNeIa~-~P, the ejecta for long living SNeIa~-~P as well SNeIa~-~T events we designate by `T', the quantities $R^{\rm cc,\,P,\,T}(r,t)$ are the rates of the corresponding SNe events, the representations for which are as follows: 
                $$R^{\rm cc}=\psi_H \int_{10}^{m_U}\xi(m)dm,$$ 
        $$R^{\rm P}=\gamma\psi_H\int_{\tau_m(8)}^{\tau_b}D_P(\tau)d\tau,$$ 
        $$R^{\rm T}=\zeta \int_{\tau_b}^{t}\psi_L(r,t-\tau)D_T(\tau)d\tau,$$
                here $D_{P,T}(\tau)$ are the so~-~called the Delay Time Distribution function for long living (subscript $T$) and short living prompt ($P$) SNeIa, $\xi(m)$ is the IMF by Kroupa (2002, hereafter Kroupa)\footnote{ After 2010 year, several new IMFs were published, e.g. Sollima (2019 and references therein). As it is stated in Section~5.1 of the cited paper, for the solar neighborhood there is a good agreement with the previous studies. The slight differences in IMFs will not noticeably manifest in the radial iron pattern since the rate of the disk enrichment is mainly determined by massive stars (see Romano et al. 2005 where the effects of various IMF were studied in details).} defined by numbers and normalized to unity:
                \begin {equation}
                \int_{0.01}^{100}\xi dm = 1.
                \end{equation}
                $\gamma$ and $\zeta $ are norming constants. Notice that in the lower limit of the integral for $R^{\rm cc}$ we substitute $m=10$, but not 8, because CC~SNe do not produce iron if their masses $8\simlt m\simlt10$. 
                
                Substituting the above relations into equation for $\mu_g(r,t)$ we compute numerically the evolution of this quantity and than the equation for iron synthesis (see equations (6) and (9) in MT19; notice that the turbulent diffusion enters explicitly into equation (9)). Simultaneously we derive the evolution of the low mass stellar components. For the high mass stellar population the instantaneous recycling approximation (IRA) was used in our previous papers(our experiments have shown that refusal from the IRA for short living objects does not change the results).
                
                Before discussion the results, notice that as it is seen from equation~(1) near the corotation resonance the difference $\Delta\Omega (r\rightarrow r_c) \rightarrow0$. Hence, within the corotation resonance $\psi_H\rightarrow 0$. Since the short living sources supply about 80 per cent of iron (MT19) such feature in the $\psi_H$ behaviour will be manifested as a gap in iron radial distribution in the corotation vicinity. However the turbulent mixing smoothes out the gap resulting in formation of a plateau~-~like metallicity distribution within the corotation.
                
                In figure 1 we demonstrate the evolution of iron large scale radial profile computed for the initial metallicity $[Fe/H](t=0)=-1.5$~dex.
                
                \begin{figure}
                \includegraphics [scale=0.96]{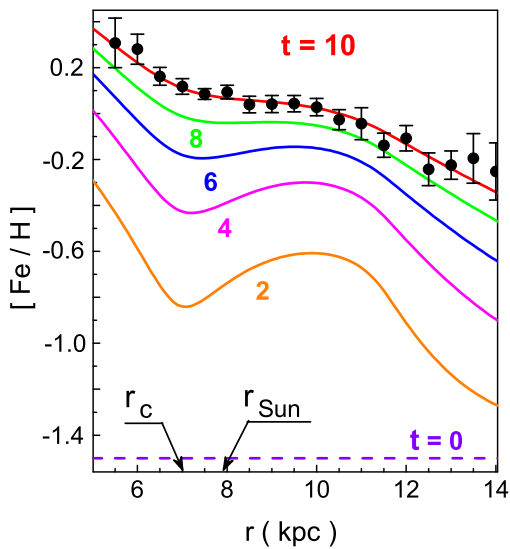}
                \caption{The evolution of radial iron abundance profiles. Numbers at lines show the corresponding moments of time in Gyr. The upper red line corresponds to $t=T_d=10$ Gyr. Filled black circles represent the observational data, averaged within the radial bins of 0.5 kpc width, the error~-~like bars are the mean metallicity scatters within the bins.} 
                \label{Fig1}
                \end{figure}

                In the figure we also show the observed iron distribution from Genovali et al. (2014, 2015) and da~Silva et al. (2016), averaged within radial bins of 0.5 kpc width.\footnote{The results, published in the cited papers, were derived by the same group of researchers, using the same method for spectral analysis of Cepheids. The metallicity scatters within the Galactic radial bins, depicted in figure~1 as error bars, are mainly associated with the natural metallicity scatter: the metallicities for the same stars do not vary from the paper to paper, cited above. Hence, we actually averaged the natural scatter of metallicities for different stars fallen into a given radial bin.} As it is seen the theoretical iron radial pattern, computed for the present epoch ($T_d=10$ Gyr is the adopted age of the thin Galactic disk), coincides well with the observed iron distribution.

                \subsection{Local MDF derivation from theoretical large scale modeling of iron radial distribution}

                The local MDF is described by the function $\Delta n_i / \Delta [Fe/H]_i$ in dependence on metallicity $[Fe/H]$, where $\Delta n_i=n_i/N$, the quantities $n_i$ and $N$ are the number of stars fallen into the $i$th metallicity bin $\Delta [Fe/H]_i$ and the total stellar number in the sample, respectively. Notice here that in the previous section we model the large scale radial iron distribution using observed iron abundances in young objects. Unlike that, in the present section, using the above elaborated theoretical model for iron synthesis, we derive the theoretical MDF over absolutely another type of objects~--~low mass stars (dwarfs), but in the close solar vicinity with the aim to compare theoretical and observed local MDFs.

        
\subsection{The standard approach to the computation of the local MDF}
                
                To compute $\Delta n_i$ we use the standard definition for the number of stars $dn$ (remind that $dn$ is a number of stars relative to the total their number in the sample or in the computation) born within the time interval from $t$ to $t+dt$ in the mass interval from $m$ to $m+dm$ (e.g. Tinsley 1980):
                \begin {equation}
                dn=\psi_L(t)\xi(m)dmdt.
                \end{equation} 
                The function $\psi_L(t)$ we adopt from the large scale radial iron distribution modeling, as it was described in the previous section (here and below we omit $r$ since the local MDF is considered at $r=r_{\odot}$). The dependences of $\psi_L$ and $[Fe/H]$ on time at the solar distance are plotted in figure~2 for two initial metallicities.                
                \begin{figure}
                \includegraphics [scale=0.99]{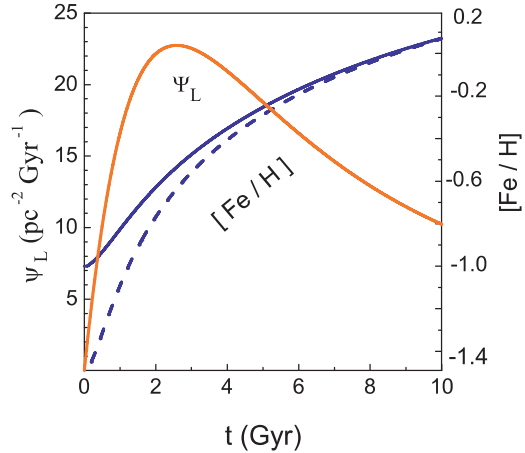}
                \caption{The temporal evolutions of $\psi_L$ for low mass stars (orange line) and metallicity $[Fe/H]$ at the solar Galactocentric distance for two initial metallicities $[Fe/H](t=0)=-1.5$ (blue dash line) and $-1.0$ dex (blue solid line).}
                \label{Fig2}
                \end{figure}

                By means of equation~(4) for known $\psi_L(t)$ we derive the number of low mass stars $\delta n_j$ which were born within a small time interval from $t_j$ to $t_j+\delta t_j$ and have survived till the present epoch:
                                
                \begin{equation}
                \delta n_j=\psi_L(t_j) \delta t_j \int_{m_1}^{m_j}\xi(m) dm,
                \end{equation}
                here $m_1$ is the least stellar mass entering the modeling, the mass $m_j$ is an upper mass of a star, that was born at $t_j$ and survived till the present epoch. For such star the following condition must be satisfied: $t_j+\tau_m(m_j)\ge T_d$.

                Further, starting at the initial metallicity, we move along the metallicity axis (see figure 2) at a sufficiently small step $\delta [Fe/H]_j=0.01$~dex and find the corresponding $t_j$ and $\delta t_j$. Then using equation~(5) we compute $\delta n_j$. Since for the MDF representation the metallicity step is $\Delta [Fe/H]_i=0.05$~dex, the quantity $\Delta n_i$ is derived by means of summation of $\delta n_j$ within the corresponding $\Delta [Fe/H]_i$.

        For comparison of the theoretical local MDF with the observed one we use the newest data from Buder et al. (2019, hereafter B19). Their parent sample includes 7,066 stars which belong both to the thin and thick disks. Since the theory, developed in MT18 and MT19, models the thin Galactic disk enrichment, we have to extract from the original B19 data the thin disk objects. For this we use the criterion described in B19 according to that the so~-~called $TD/D$ parameter for stars, which have disk like kinematics, is less than 0.5 (details see in B19 and references therein). After application of this procedure to the parent data the number of stars in the sample occurs to be 6,672.
        
        From the remaining stars we remove those ones which ages are greater than $T_d=10$~Gyr. During this step we take into account the errors in the stellar ages given in B19 for each star. To illustrate the influence of the random errors in stellar ages on their distribution on masses we perform numerical experiments supposing that the age of a star (from the remaining 6,672 ones) is randomly distributed according to Gaussian law at a spread equal to the given in B19 sample standard deviation for each star relative to its mean value. After this procedure, repeated 50 times, in different computation sets the number of objects varied from $\sim6,200$ to $\sim6,500$.

        Further to demonstrate the combined effect of stellar errors in metallicities and in their ages on the observed MDF we additionally randomly `noise' mean $[Fe/H]$ for an individual star (from the remaining sample), consider the estimated by B19 error in metallicity $\sim0.075$~dex as the spread (in the experiments the generated random sequences for stellar ages and metallicities are statistically independent). The resulting 50 disturbed observed MDFs are shown in figure~3.

        For illustration of the problem, in figure~3 we show the theoretical MDFs, computed for several input parameters and superimposed on the observed 50 noised MDFs. The theoretical MDF were computed for several input parameters: i)~the initial metallicity $[Fe/H](t=0)=-1.5$~dex and $m_1=0.01$ (according to Kroupa this mass is the least stellar mass, see equation~(3)); ii)~$[Fe/H](t=0)=-1.0$~dex and  $m_1=0.01$; iii)~ $[Fe/H](t=0)=-1.0$~dex but $m_1=0.49$ (the last value corresponds to the least stellar mass in B19 sample, this star being the only one in the parent B19 sample which mass is less than 0.5~M$_{\odot}$; in the abov computations the upper stellar mass is $m_2=2.1$ as in B19 data). 
        
        Our experiments with the correction factors, used by Haywood, do not improve the fit of the theoretical and observed MDFs within the standard approach.

                \begin{figure}
                \includegraphics [scale=0.99]{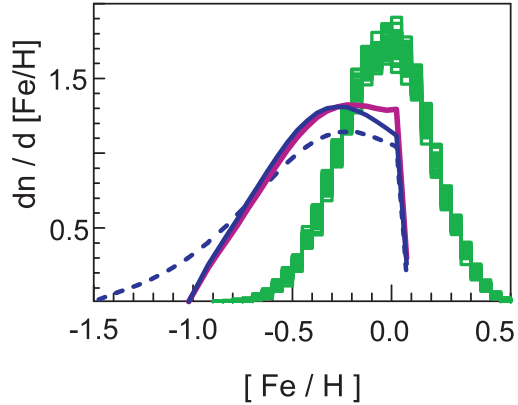}
                \caption{{\it Green histograms}: the noised 50 observed MDFs derived using B19 data. {\it Blue dash line}: the theoretical MDF computed for the initial metallicity $-1.5$~dex and the lowest stellar mass $m_1=0.01$. {\it Blue solid line}: The same as previous but for the initial metallicity $-1.0$~dex. {\it Magenta line}: The same as previous but for $m_1=0.49$. The theoretical MDFs were computed using Kroupa IMF. The bin widths $\Delta [Fe/H] =0.05$ dex. Hereafter the quantity $dn$ is defined as the number of stars in a bin relative to the total their number in the sample (see text for details). All distributions are normalized to unity (i.e., the squares under the curves are equal to 1).}
                \label{Fig3}
                \end{figure}
                                                
                We also notice the abrupt fall of the theoretical distribution at $[Fe/H]\approx0.1$~dex. It is expected since at present time the abundance must be close to the mean metallicity observed in the local solar vicinity (see figure~2). Nevertheless the absence of stars in the metallicity range $[Fe/H]\simgt0.1$~dex also requires an additional consideration (see below).

        
                \subsection{Modification of IMF}
                        
                What is a cause of the divergence between the theoretical and observed MDFs demonstrated in figure 3? Is it possible that the observed sample is not a representative one? Or the theory has some inherent flaw?
        
To explain the observed metallicity distribution function over low mass stars we must bear in mind many attendant circumstances like closeness of the solar vicinity to the peculiar region in the Galactic disk~--~the corotation resonance (see figure~1). Due to that the enrichment of the solar neighbourhood by iron is an atypical with respect to surroundings. Moreover, as it was shown by Mishurov \& Acharova (2011) low mass stars, born near the corotation radius, will be scattered over a large portion of the Galactic disk during several billion years under the influence of spiral arms, etc. However these effects were not considered in MT18 and MT19.
        
        On the other hand, owing to long life times, comparable to the Galactic disk age, significant portion of low mass stars survives up to the present epoch and can contribute to the metal~-~poor tail of {\it the theoretical} MDF. But if low mass stars are missed in observations (due to their low luminosities) we derive the discrepancy comparing the observed and theoretical MDFs since the number of low mass stars will be overestimated in the theoretical MDF with respect to the observed sample.
        
        To estimate the portion of missed stars in observed sample we adopt the stellar density in the solar vicinity $\sim40$~M$_{\odot}$ pc$^{-2}$ (Haywood et al. 1997) and a typical stellar mass $\sim1$. If an observed sample (like the B19 one) contains $\sim7,000$ stars within 100 pc around the Sun their portion occurs to be less than 1~per cent relative to the real stellar number in the solar vicinity. At first sight, the above observed sample has too few stars. Nevertheless this is the best up to date sample which contains both the metallicities and stellar masses.\footnote{The great progress in this area is foreshadowed due to Gaia mission. In particular, an unprecedented large number of stars (more than 120,000) were used for derivation of IMF by Sollima (2019). However, the data on stellar metallicity for these objects are absent in this study. That is why we are forced to work with B19 sample.} But, before fitting the theoretical MDF to the observed one, we have to adjust the stellar distributions over their masses.

        How to adjust the theoretical and observed stellar distributions on masses? To provide an answer this question, in figure 4 we display the observed stellar distribution on masses from B19 data (50 randomly noised histograms as it was described in the previous Section). From the figure it is obvious that, in the mass range $m\simlt1$ the fraction of stars in the observed sample decreases with stellar mass decrease whereas the IMF by Kroupa increases with stellar mass decrease! 
  \begin{figure}
  \includegraphics [scale=0.96]{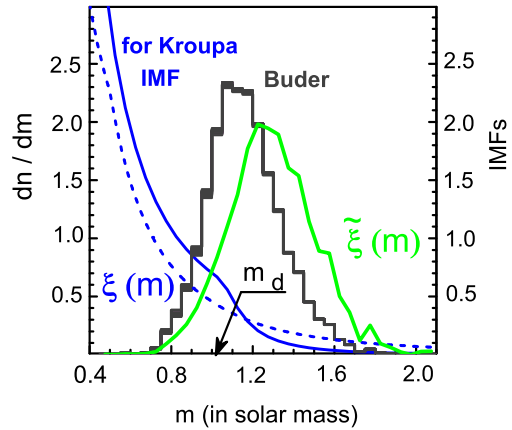}
   \caption{{\it Dark grey histograms}: the 50 noised observed stellar distributions on masses, $(dn/dm)_{obs}$ for B19 sample. {\it Blue dash line}: the IMF by Kroupa. {\it Blue solid line}: the theoretical $dn/dm$ computed by means of equation~(6) for Kroupa IMF. {\it Green solid line}: the modified IMF (mIMF) computed by means of equation~(7). The distributions are normalized to unity. The quantity $m_d$ is the mass of a star which life time is equal to the age of the Galactic disk.} 
                \label{Fig4}
                \end{figure}
        
        In figure~4 we additionally plot the theoretical stellar distribution on masses, $dn/dm$, derived by means of the following equation:
        \begin{equation}
                \frac{dn}{dm}=\xi(m)\int_{T_d-\tau_m(m)}^{T_d}\psi_L(t)dt,
                \end{equation}
using Kroupa IMF. The lower limit in the integral differs from zero if $\tau_m(m)<T_d$. This means that we are only interested in the stars which have survived till our epoch (see the explanation for $m_j$ above).

        Comparison of the behaviour of this theoretical distribution with the observed one demonstrates their drastic discrepancy. It is obvious that, strongly excessive number of low mass (correspondingly long living) stars will excessively increase the metal~-~poor tail in the theoretical MDF with respect to the observed one.
        
        To adjust the theoretical stellar distribution on their masses to the corresponding observed distribution we propose to invert the equation~(6) supposing that the stellar distribution on masses is known from the observed by  B19 (denote it as $(dn/dm)_{obs}$) and the SFR we take from MT18 and MT19. The representation for the mIMF (denote it as $\tilde{\xi}(m)$) is as follows:
                                
                \begin{equation}
                \tilde{\xi}(m)=\frac{(dn/dm)_{obs}}{\int_{T_d-\tau_m(m)}^{T_d}\psi_L(t)dt}.
                \end{equation}
                The derived mIMF we also display in figure~4. As it was expected its behavior absolutely differs from the standard IMF by Kroupa. 
        
        Before discussion the effects due to application of the modified IMF let us summarize the written above. At first step, in the modelling of galactic nucleosynthesis the IMF by Kroupa must be used since the observed values like gaseous and stellar densities, etc., at present epoch have to be derived. However, to extract from this modeling the theoretical metallicity distribution function over low mass stars, the excessive number of them (relative to their number in observed sample) must be removed from the models, so as the theoretical stellar distribution on masses fits the used observational sample (in our case the B19 data).

                
                \subsection{Effects of modified IMF, stellar wandering and random scatter in metallicity}

                Substituting the above mIMF instead of Kroupa one into equation~(5) we compute a new theoretical MDF which is shown in figure~5. As it is seen the structure of the theoretical MDF now has been significantly changed relative to the ones plotted in figure~3.
                                
                \begin{figure}
                \includegraphics [scale=0.99]{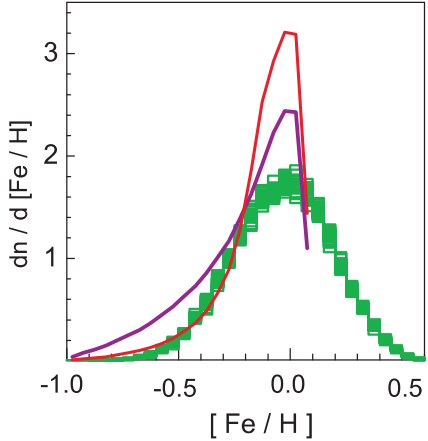}
                \caption{{\it Green histograms}: the noised observed MDFs as in figure~3. {\it Purple line}: the theoretical MDF derived by means of mIMF. {\it Red line}: The same as above but additionally corrected using SM91 correction factor numbered as 3 in Haywood (2006). The both theoretical distributions were computed for initial $[Fe/H]=-1.0$~dex.} 
                \label{Fig5}
                \end{figure}
                                
                Our subsequent step is to apply correction factors to the above new theoretical MDF. These factors take into account the stellar wandering mainly across the Galactic disk. They were analyzed by Haywood (2006). The strongest correction effect will bring the factor derived by SL91 and numbered as 3 in Haywood (2006). The result of its implication is also plotted in figure~5. Now the match of the theory with observations in the metallicity range $[Fe/H]\simlt-0.2$~dex proves be very good.
                                
        The next step is to take into account the effect of random scatter in metallicity which presents in the data both due to intrinsic (i.e. natural) causes and errors in abundance measurements.\footnote{Remind that the theory of the Galactic chemical evolution, developed in the framework of fluid approach, deals with mean values at a given point in the disk.} For this we introduce the randomization into the above theoretical MDF by analogy with Haywood (2006, see also other papers of this author), the random scatters being distributed according to the Gaussian law. The resulting `noised' theoretical MDFs, computed for two standard deviations (0.15~dex and 0.18~dex), are shown in figure~6. The coincidence of the theoretical MDF with the observed one is fairly well.
        
Nevertheless we notice that the theoretical MDFs are slightly shifted ($\simlt -0.1$~dex) relative to the observed one. This shift maybe associated with the atypical location of the solar vicinity in the Galactic disk (by the way, Haywood 2014 also suggested that location of the solar vicinity maybe an atypical). Indeed, as it is seen from figure~1, the solar vicinity has spent too much time (from $t=0$ and till to $t\sim7$ Gyr) within the region of decreased ($\sim0.1$ dex) iron abundance. The physical mechanism for the shift may also be associated with strong stellar radial wondering under influence of gravitational field from Galactic spiral density waves, particularly due to closeness of that stars to the corotation resonance (Mishurov \& Acharova 2011). But in our previous papers this effect was not taken into consideration. We plan to study it in future works.
                                                        
                \begin{figure}
                \includegraphics [scale=0.98]{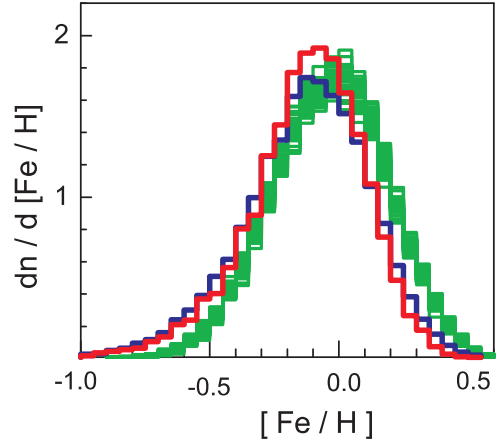}
                \caption{{\it Green histograms}: The same as in the previous figure. {\it Red histogram}: The theoretical MDF from figure~5 (plotted by the red line) randomized using the metallicity scatter with the spread 0.15~dex. {\it Blue histogram}: the same as previous but for the spread 0.18~dex. For details see the text.}
                \label{Fig6}
                \end{figure}

                                
                                \section{Conclusions}
                                
                In MT19 we refine a model, previously proposed by AGMK13, for explanation of the observed large scale wriggling iron radial pattern along the Galactic disk. Such nontrivial radial iron distribution in the disk was explained by means of combined effect of the corotation resonance and turbulent diffusion of iron. The influence of the corotation resonance on iron Galactic enrichment rate is associated with the strong decrease of short living ($\simlt100$ Myr) star formation rate in the corotation vicinity (remind that, as it was shown in MT19 these stars supply $\sim80$ per cent of iron to the Galactic disk). Such decrease results in the formation of the gap in iron abundance near the resonance that is subsequently smoothed out due to turbulent diffusion and leads to formation of a plateau in the radial iron pattern in the solar vicinity (see figure~1).
                
                As the observational data, MT19 use Cepheids. Being bright objects, they are seen at large accurately measured distances, but as very young stars (their ages are $\simlt100$ Myr, AMK12) Cepheids delineate the iron radial abundance pattern at present epoch.
                                
                Can the model by MT19 be reconciled with the metallicity distribution function observed over absolutely another type of objects like low mass long living dwarf stars in the local solar vicinity?

                Our studies show that, direct derivation of the local MDF for low mass stars from the model by MT19 failed to reconcile it with the observed MDF (see figure~3). This discrepancy of the theory with observations we associate with use of the standard IMF: as it is shown in figure~4 the use of the standard IMF leads to the theoretical stellar distribution on masses which absolutely differs from the observed by B19: in the theoretical distribution low mass and long living stars are overestimated relative to the observed one. Hence in this case we will expectedly derive the excessive number of metal~-~poor stars.
        
        To improve the situation we propose to transform IMF so as to extract from the theory by MT19 a portion of stars (remind that the portion of stars in the observed sample is $\simlt1$ per cent of their number in the Galactic disk, see Section 2.4) which distribution on masses is close to the observed one. As a consequence, the theoretical MDF now occurs to be closer to the observed one (see figure~5).
        
        After that, applying the corrections proposed in the series of papers by Haywood we achieve fairly well coincidence of the theoretical and observed MDFs (see figure~6). Slight shift between theoretical and observed MDFs maybe associated with stellar radial wondering under the influence of spiral arms revealed by Mishurov \& Acharova (2011).

                                \section*{Acknowledgements}
                                
        Authors thank an anonymous referee for their insightful comments and many useful suggestions which encouraged us to revised our modelling algorithm. Yu.N.M. acknowledges financial support from the Southern Federal University of Ministry of Science and Higher Education of Russian Federation.

                \end{document}